# Comparing People with Bibliometrics

Michael J. Kurtz
Harvard-Smithsonian Center for Astrophysics
ORCiD: 0000-0002-6949-0090

## Abstract

Bibliometric indicators, citation counts and/or download counts are increasingly being used to inform personnel decisions such as hiring or promotions.  These statistics are very often misused.  Here we provide a guide to the factors which should be considered when using these so-called quantitative measures to evaluate people.  Rules of thumb are given for when begin to use bibliometric measures when comparing otherwise similar candidates.

## Some things you should know when using bibliometrics to compare people

This is a serious business.  Careers can be greatly affected by the decisions which flow from bibliometric evaluations. [1,2]

It is important to know how accurate these measures are.  Measures such as citation or download counts are not important in themselves, they are proxies for evaluating the (research) capabilities of an individual. [1,3,4]

While the relative completeness and accuracy of some sources of bibliometric information are frequently discussed, this is not particularly important to comparisons of people.  The statistical scatter, even with "perfect" input data in these measures is quite large, and puts limits on any attempt to use the data to evaluate people. [1]

Before using citations or downloads to compare individuals these measures must be adjusted to account for the known systematic errors due to age, discipline, and co-authorship. [5]

The total number of citations a person's work receives increases approximately quadratically with the "age" of that individual, where "age" means time as an active researcher. [5,6]  The correction is that the square root of the number of citations, divided by "age" is an approximate constant for persons with uniform productivity.  For younger researchers determining the effective "age" is problematic, and makes this correction uncertain. [1]

Different disciplines, and subdisciplines, have different citation cultures, citation counts cannot be directly compared without correcting for  this. [5,7]

Multi-author papers are now the norm. Bibliometric measures must take the degree of co-authorship into account. [5,8,9]

For comparisons with individuals measured at different times one must account for the steady increase in the number of papers, which double about every 15 years [10,11], and in the number of references per paper, which is experiencing a similar increase [12,13].

The TORI statistic [5], which normalizes citations by both the number of authors on the cited document and the number of references in the citing document, is intended to remove the effects of co-authorship and citation culture. The RIQ statistic [5] is intended to remove the effects of age. Both are available on the ADS metrics pages.

Downloads, or reads, are also a useful bibliometric indicator [4,9,14]. Downloads by researchers have very similar properties to, and can predict citations [11,14,15]. Great care must be taken when using download information, however, as downloads by persons who are not researchers have very different properties, and can dominate the statistic [4,7,13].

The ADS metrics pages show the Read10 statistic [1,9], which is the yearly sum of co-author number normalized reads by scientists via ADS of papers written in the preceding ten years. Because the use of internet based services has grown rapidly in the recent past care must be taken when comparing download counts from different time frames.

Using samples and measures designed to minimize the systematic errors in the measurements we can determine the intrinsic scatter in the use of bibliometric techniques to evaluate people. We do this by comparing different measures of exactly the same sets of papers by exactly the same people. Because the actual impact of these sets of papers is identical (because they are exactly the same) the differences in impact as measured by different techniques must be due to the intrinsic measurement error in the techniques. Complete details are in [1].

Because citations and reads of people form a log-normal distribution [1,17] the error is in the logarithm of the measurements. For the comparisons using active mid-career astronomers the one standard deviation error is 0.17 dex, a multiplicative factor of 1.48. [1]

Citations (and downloads, by extension) are also used to predict performance; essentially this is why they are used in hiring decisions. For individuals five years past the PhD, about the time when assistant professor positions are granted, even using an *a posteriori* selected sample, substantially more prescient than any contemporaneously selected sample could be, the one standard deviation error is 0.395 dex, a multiplicative factor of 2.5. [1]

## Conclusions

Bibliometric measures, citation and download counts, have been shown to be useful and accurate in a number of areas.  When aggregated over large entities, such as universities [13,18], countries [13,29], or journals [20] bibliometric measures can provide accurate measures of the quantity and quality of research.  Additionally, persons with very high citation counts are, almost always, very well regarded researchers.  This means that citations can be accurately used to recognize these individuals, and predict high honors [3].

These successes have led some to use bibliometric measures to evaluate people.  In most cases this constitutes a misuse of the statistics.  Decision aids such as citation or download statistics may well be able to identify most of the top people in a field, but they have almost no power in making the more common 41st chair [21] type of personnel decision: deciding between a set of similarly qualified candidates.

Because in any real instance the various corrections for age, co-authorship, subfield, etc will not be as clear as in the near ideal case analyzed in [1] it is reasonable to require that, when comparing two individuals, the difference in their citation or download measures be greater than two standard deviations (95% confidence) before ascribing any difference to them on the basis of those counts.  This would be about a factor of two when comparing the current status of mid-career scientists, and about a factor of five when predicting the future productivity of possible assistant professor hires.  Outside of these bounds one would want to understand why the counts were so different.